\documentclass{article}
\usepackage{spconf,amsmath,graphicx}

\usepackage{enumitem}
\setlist{nosep, leftmargin=14pt}

\usepackage{mwe} % to get dummy images
\usepackage{hyperref} 
\title{multi-modality microscopy image style transfer for nuclei segmentation}
\twoauthors
 {Ye Liu}
	{Technical University Munich
	}
 {Sophia J. Wagner\sthanks{The second author was supported by the Munich School for Data Science (MUDS).}, Tingying Peng}
	{Helmholtz Zentrum M\"unchen, Helmholtz AI
	}
\begin{document}
%\ninept
%
\maketitle
\begin{abstract}
Annotating microscopy images for nuclei segmentation is laborious and time-consuming. To leverage the few existing annotations, also across multiple modalities, we propose a novel microscopy-style augmentation technique based on a generative adversarial network (GAN). Unlike other style transfer methods, it can not only deal with different cell assay types and lighting conditions, but also with different imaging modalities, such as bright-field and fluorescence microscopy. Using disentangled representations for content and style, we can preserve the structure of the original image while altering its style during augmentation. We evaluate our data augmentation on the 2018 Data Science Bowl dataset consisting of various cell assays, lighting conditions, and imaging modalities. With our style augmentation, the segmentation accuracy of the two top-ranked Mask R-CNN-based nuclei segmentation algorithms in the competition increases significantly. Thus, our augmentation technique renders the downstream task more robust to the test data heterogeneity and helps counteract class imbalance without resampling of minority classes.
\end{abstract}
\section{Introduction}
\label{sec:intro}

The evaluation of cell-level features is a key task in the histopathological workflow. Features, such as nuclei shape and distribution, are used to determine cell, tissue, and cancer types and are therefore relevant for the clinical diagnosis of cancer, e.g., for cancer identification, grading, and prognosis \cite{gurcan2009histopathological}. Deep learning serves as a promising tool to quantify these features in an automated manner \cite{2019}.
However, the performance of these models depends heavily on the quality and quantity of training data since they require accurate and time-consuming segmentation masks. To tackle this, the Kaggle competition 2018 Data Science Bowl (DSB)\footnote{\url{https://www.kaggle.com/c/data-science-bowl-2018}} was initiated to find segmentation algorithms that are robust and effective in various microscopy set-ups, cell-lines, and, most significantly, types of light microscopy. Among 3,891 submissions during the challenge, the top-performing methods were mostly based on the most common segmentation network architectures, Mask R-CNN \cite{He2017} and U-Net \cite{Ronneberger2015}. The most important factors influencing the competition ranking were the amount of training data\footnote{Some teams also used private training data aside from the data provided by the challenge organizer}, complex pre-processing and post-processing, and ensemble learning. Yet, a common problem that encumbers the performance is dataset bias: while segmenting fluorescent images, which make up most of the training set, is relatively easy and accurate, segmenting less-represented image types, such as bright-field images of stained tissue sections, is difficult and inaccurate \cite{Caicedo2019}.

In this paper, we propose to facilitate network training by a GAN-based style transfer data augmentation technique as has been shown to be very effective for histological images \cite{HistAuGAN}. By synthesizing less-represented image types from well-represented ones, our style augmentation can increase the amount of images of minority types in the training set. More specifically, we use disentangled representations in the style transfer, i.e., we decompose the input into its content and style representations and only alter the style without changing the content. Since the structure is preserved, the style augmented images have the same nuclei locations and shapes as the original image. Therefore, our augmentation technique enables the segmentation network to see a multitude of styles already during training, rendering it more robust to the heterogeneity in the test data.

\section{Method}
\label{sec:pagestyle}
\begin{figure*}[htb]
    \centering
    \centerline{\includegraphics[width=16cm]{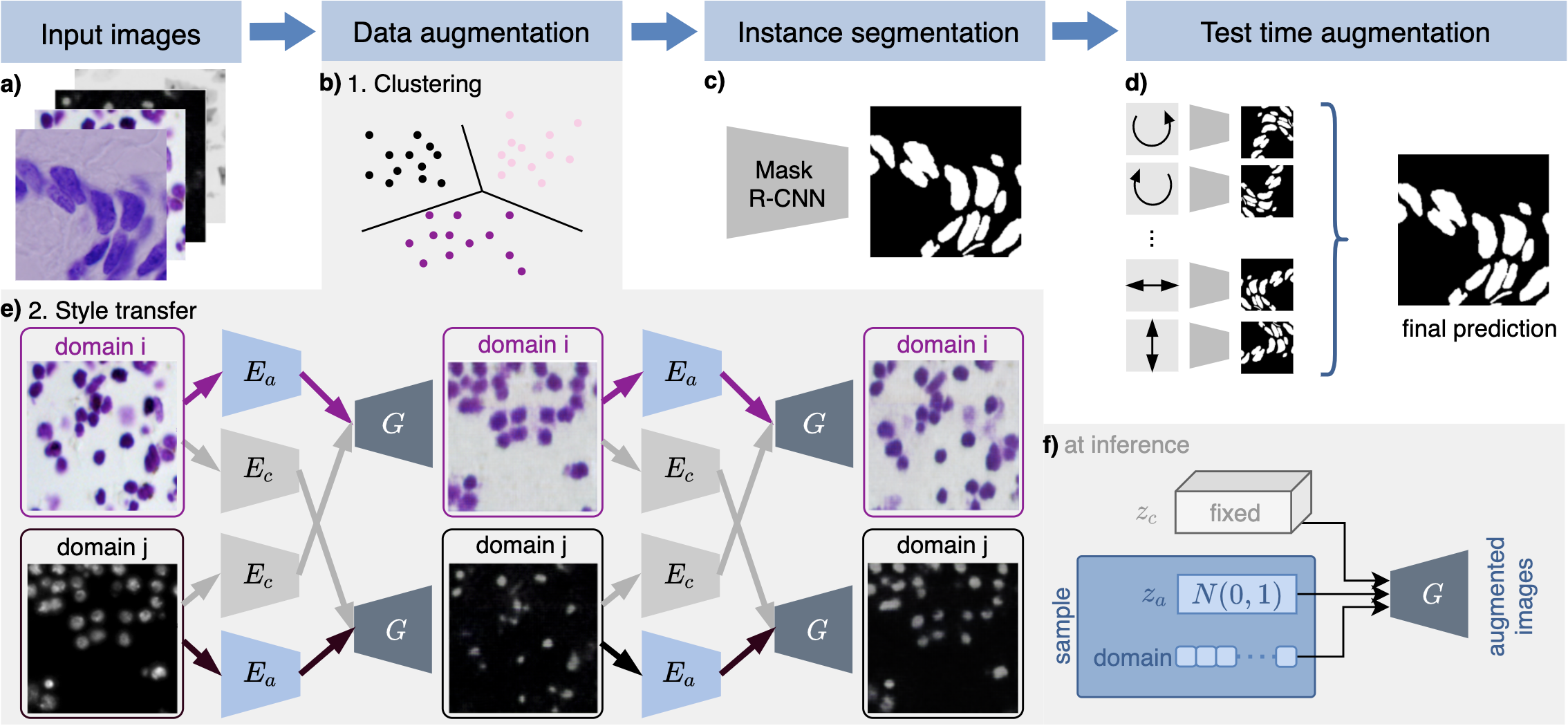}}
    \caption{Overview of the nuclei segmentation pipeline with multi-modality style transfer for data augmentation.}
    \label{fig:workflow}
\end{figure*}

Our nuclei segmentation workflow consists of data augmentation, split into clustering and style transfer, training an instance segmentation network and, finally, evaluating it with test time augmentation as shown in Figure \ref{fig:workflow}.
In the following, we detail each of the steps.

\subsection{Dataset description}
\begin{figure*}[htb]
\begin{minipage}[b]{1.0\linewidth}
  \centering
  \centerline{\includegraphics[width=17cm]{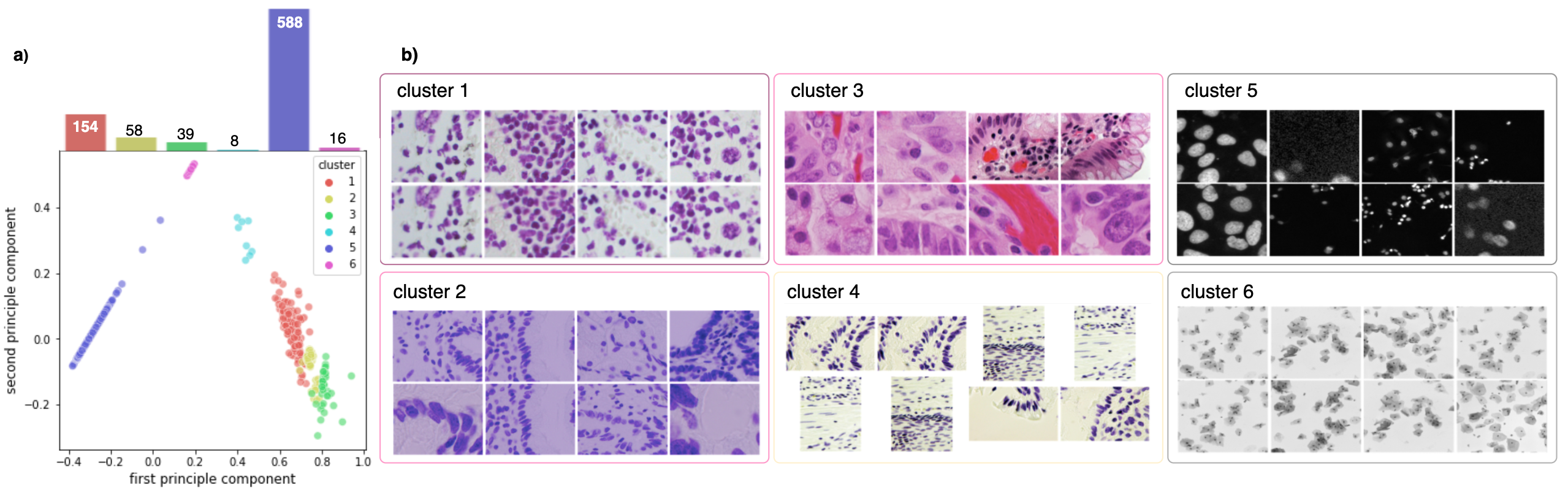}}
\end{minipage}
\caption{(a) PCA decomposition of hue saturation values of all training images and their class distribution. Each point represents one image, colored according to the corresponding cluster. (b) Exemplary images for each cluster.}
\label{fig:cluster}
\end{figure*}

The dataset was provided by the competition organizers and is publicly available in the Broad Bioimage Benchmark Collection with access number BBBC038\footnote{\url{https://bbbc.broadinstitute.org/BBBC038}}. It is divided into three sets: a training set of 670 images with nuclei segmentation masks provided (29,464 nuclei), a first-stage test set of 65 images (4,152 nuclei), and a second-stage test that includes 3,019 images (37,333 nuclei). The evaluation of nuclei segmentation for the annotated first-stage test set was disclosed later by the organizers. However, the official evaluation and ranking of a segmentation algorithm are based on the performance of the second-stage test set, which hides the annotated nuclei masks from the participants. Both training and test images are very diverse in terms of lighting conditions, cell lines, nuclei densities, and microscopy imaging modalities (with a few examples shown in Figure \ref{fig:workflow}a and \ref{fig:cluster}). It is worth noting that the test set is not only substantially greater in size than the training set, but also includes novel image modalities that have never been seen during training, rendering the segmentation task extremely challenging.  

\subsection{Data augmentation step I: clustering training images into different modalities}
Since the dataset does not contain any metadata that can be used to determine the image modalities, we first divide our training images into multiple domains to perform image style transfer. Each cluster should ideally correspond to a different modality.  As a result, we used the classic K-means algorithm to divide these images into six distinct clusters based on their hue saturation value (HSV). Each cluster should ideally reflect a modality that constitutes a distinct domain. We also observed that the fluorescent images are typically dark and have low contrast. So we use the Contrast Limited Adaptive Histogram Equalization (CLAHE)\footnote{\url{https://scikit-image.org/docs/dev/auto_examples/color_exposure/plot_equalize.html}} for contrast enhancement. For each section of an image, CLAHE computes a contrast histogram and adjusts the local contrast accordingly if that section is darker or brighter than the rest of the image.
\subsection{Data augmentation step II: multi-modality style transfer}
As shown in Figure \ref{fig:workflow}e, for two exemplary images from different domains, the modality transfer model based on \cite{DRIT} first decomposes each image into a domain-invariant content using a content encoder $E_{c}$, and a domain-specific attribute using an attribute encoder $E_{a}$, respectively. The content encoding and the attribute encoding, together with the domain information, can then be used to generate a synthetic image, where the structure is preserved. To create various synthetic images as augmentation during training of the segmentation network, we sample attribute and domain vector randomly, while leaving the content-encoding fixed (see Figure \ref{fig:workflow}f). We apply this augmentation technique randomly to half of the training images additionally to standard augmentations. 

The content space is assumed to be shared across the domains. Therefore, during training, the content vector $z_c$ is exchanged between the modalities. To compute the cross-cycle consistency loss, this is repeated to reconstruct the original image since we trained without paired images. The final objective function is given by
\begin{eqnarray*}
L_\text{total} &=& w_{cc}L_{cc} + w_cL_c + w_dL_d + w_\text{recon}L_\text{recon} \\
&&+ w_\text{latent}L_\text{latent} + w_{KL}L_{KL},
\end{eqnarray*}	
where $L_{cc}$ is the cycle-consistency loss, $L_c$ and $L_d$ are adversarial losses for the content and the attribute encoder, $L_\text{recon}$ is an $L_1$-loss for image reconstruction (synthesizing an image using its own content and attribute), $L_\text{latent}$ is an $L_1$-loss for latent space reconstruction, and $L_{KL}$ enforces the latent attribute space to be distributed according to the standard normal distribution. Please refer to \cite{DRIT} for a detailed explanation of each loss. 

Since the content encoding is fixed during augmentation, the augmented image has the same nuclei location and shape as the input image and thus inherits the nuclei segmentation mask from the original image. This is a key difference between our approach and a common cycleGAN-based image style transfer \cite{2017cycleGAN}, as there is no guarantee of content invariance after the cycleGAN transfer.

\subsection{Mask R-CNN for instance segmentation}
For nuclei segmentation, we used two implementations published in the competition leaderboard\footnote{\url{https://github.com/Lopezurrutia/DSB_2018}, 

\url{https://github.com/mirzaevinom/data_science_bowl_2018}} as baselines. Both teams use  Mask R-CNN \cite{He2017} for segmentation. The implementations differ in the image pre-processing and preparation of the network training. We choose these two methods because of their publicly accessible code, sufficient documentation, excellent performance (both methods were ranked among the top-5 in the leaderboard), as well as relatively simple network training (only one model needs to be trained, by contrast, the top-1 approach trained 32 U-nets, which would consume considerably more training time and resources). Notably, the second approach developed by the team mirzaevinom used additional data beyond the data provided by the organizer. However, we only used the annotated training data provided by the organizer to train our model so our model performance differs slightly from their results reported in the leaderboard. For each method, we train two models, one with and one without multi-modality style transfer. 

\subsection{Test time augmentation}
In the inference stage, we use test time augmentation (TTA). TTA uses simple geometric augmentation on test images, e.g., rotating, flipping, color jittering, and image rescaling, and aggregates the model predictions on these augmented images. Thereby, TTA improves the robustness of Mask R-CNN without any additional training cost, leading to an increase in segmentation accuracy when compared to prediction on the original image \cite{Moshkov2020}.

\section{Results and Discussion}
\label{sec:typestyle}

\subsection{Clustering into modalities}
Figure \ref{fig:cluster}b shows that the training set can be clustered into six sets that correspond indeed to the different modalities. These clusters can also be seen in the low-dimensional principal component analysis (PCA) embedding of the training data visualized in Figure \ref{fig:cluster}a. We observe that the training set is highly imbalanced: over $75\%$ of all training images are fluorescent images, whilst the remaining bright-field images are grouped into five distinct, partly very small clusters. 

\subsection{Improved segmentation performance by style transfer augmentation }
\begin{figure}[t]
  \centering
  \centerline{\includegraphics[width=\linewidth]{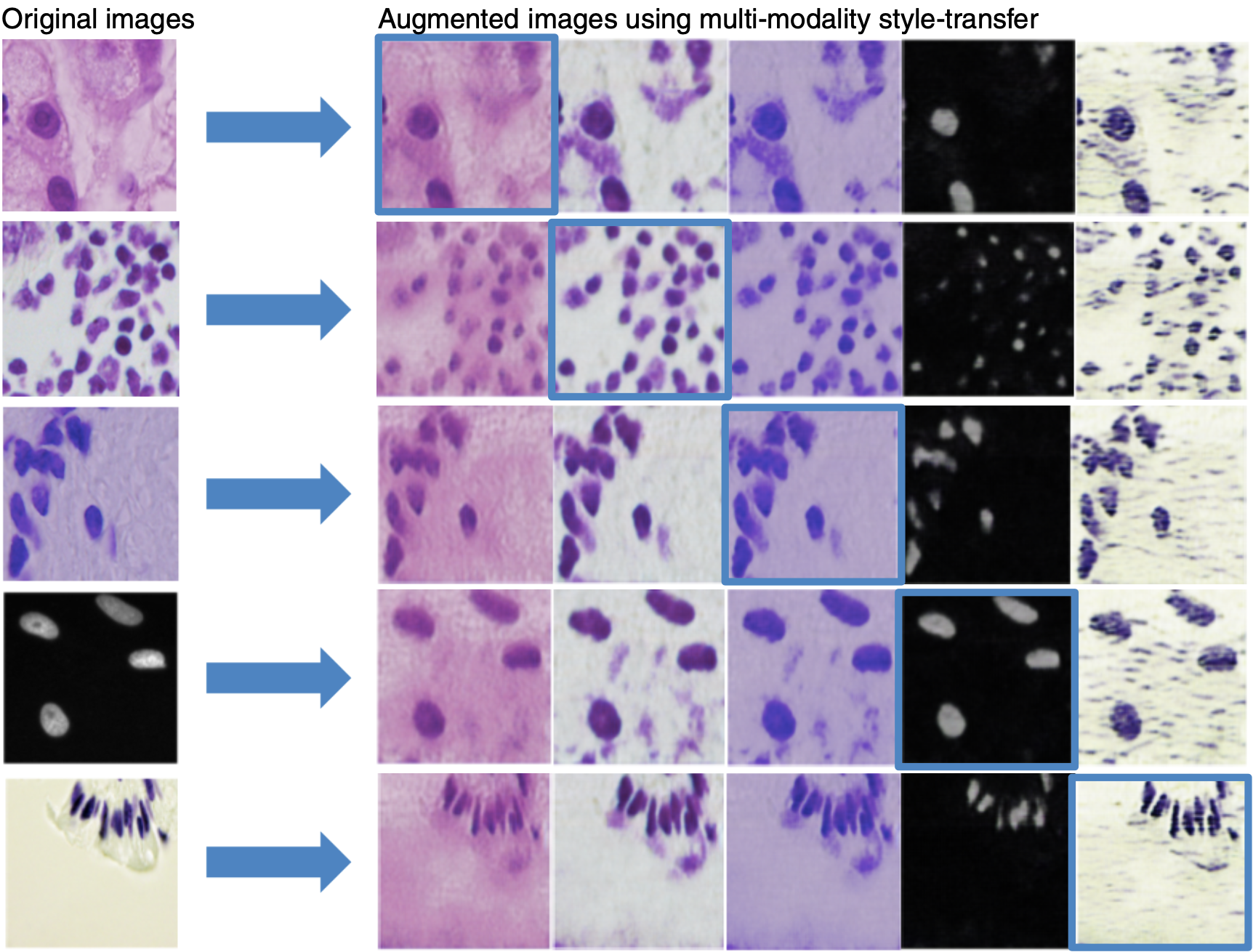}}
  \caption{Examples of the outcome of multi-modality style transfer. First column: original images. Remaining columns: result of each domain's multi-modality style transfer with self-reconstruction on the diagonal, highlighted in blue boxes.}
  \label{fig:style_transfer}
\end{figure}

Figure \ref{fig:style_transfer} shows a few examples of images created by our multi-modality style transfer GAN from one domain to the others. Visual assessment reveals that our generated images resemble real ones, implying that they are well suited to the augmentation task.  

\begin{figure}[t]
  \centering
  \centerline{\includegraphics[width=1\linewidth, angle=0]{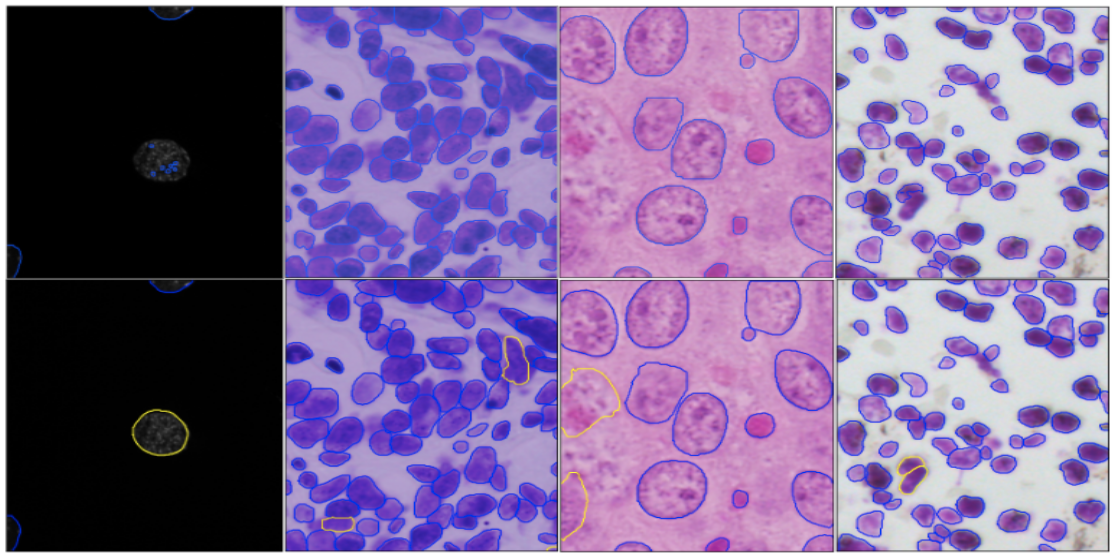}}
  \caption{Segmentation results without (first row) and with (second row) our augmentation. Predicted mask contours in common are shown in blue. Yellow contours in the second row show the improvements compared to the first row.}
  \label{fig:seg}
\end{figure}

We quantify the add-on value of our proposed augmentation method by training models with and without augmentation of the multi-modality style transfer GAN, trained with only the training data set. The final evaluation score is based on the Intersection-over-Union (IoU) metric, determined by submitting our segmentation results of the second-stage test dataset to the Kaggle competition\footnote{\url{https://www.kaggle.com/c/data-science-bowl-2018/submit}}. As shown in Table \ref{fig:result}, including style transfer augmentation in the method increased the nuclei segmentation accuracy from $53.2\%$ to $60.9\%$ for the method I and $59.9\%$ to $61.3\%$ for method II, respectively. Notably, a score around $61\%$ was ranked among the top-5 and almost only achieved by using additional datasets. In Figure \ref{fig:seg}, we can observe that our augmentation technique using the multi-modality style transfer GAN improves the segmentation results, especially in less-represented image modalities such as bright-field images of stained tissue sections and fluorescence images of large nuclei (most fluorescent images in the training set contain only small nuclei). 

In addition to the above two baseline methods, we also quote the result from the team BIOMAGic ($57.0\%$ IoU), as it was the only one to use style transfer among the top-25 submissions during the competition. Unlike our style transfer between multiple microscopy image modalities, BIOMAGic (and its successor nucleAIzer \cite{Hollandi2020}) perform style transfer between images and their segmentation masks, which can be trained only by annotated images. By contrast, our method can be trained in an unsupervised fashion using images without annotated segmentation masks.  In particular, their style transfer approach required training 134 networks for fine-grained image clusters for the same competition, whereas our approach only required the training of one network. 

% It is worth mentioning that we didn't use any external data, while the team Inom Mirzaev did. Besides, our methods are based entirely on publicly available codes.
\begin{table}
\begin{center}
%\scalebox{0.9}{
\begin{tabular}{ l|c } 
\hline  
 Methods & IoU score\\
\hline
BIOMAGic &  0.570\\ \hline
Deep Retina w/o our augmentation &  0.532   \\ 
Deep Retina w/ our augmentation& \textbf{0.609}  \\ \hline 
Inom Mirzaev w/o our augmentation& 0.599  \\ 
Inom Mirzaev w/ our augmentation& \textbf{0.613}  \\ 
\hline  
\end{tabular}
\caption{IoU scores of different segmentation methods.} 
\label{fig:result}
\end{center}
\end{table}

\begin{figure}[htb]
\begin{minipage}[b]{\linewidth}
  \centerline{\includegraphics[width=4cm]{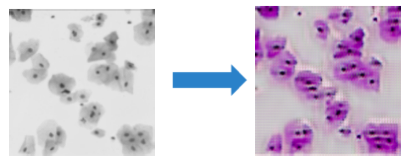}} \medskip
\end{minipage}
\caption{The style transfer GAN fails when images at two modalities contain different content.}
\label{fig:discussion}
\end{figure}

Although our approach is effective in transferring style across the majority of modalities, it also has limitations. When using our style transfer GAN from image modalities that only show nuclei to another modality where both nuclei and cytoplasm are visible, unrealistic images are generated as demonstrated in Figure \ref{fig:discussion}. This is because that these two modalities do not only differ in styles but also in content, hence the content and style cannot be efficiently disentangled.

\section{Conclusion}

In summary, we developed an augmentation technique using a multi-modality style transfer GAN to transfer microscopy nuclei images from one modality to another. During training a Mask R-CNN for nuclei segmentation, this augmentation strategy facilitates the training by increasing the diversity of the training images, hence making it more robust to the test data heterogeneity and resulting in better segmentation accuracy.

\section{Compliance with Ethical Standards}
This research study was conducted retrospectively using human subject data made available in the Broad Bioimage Benchmark Colletction with accession number BBBC038\footnote{\url{https://bbbc.broadinstitute.org/BBBC038}}. Ethical approval was not required as confirmed by the license attached with the open-access data.

\vfill
\pagebreak

% Below is an example of how to insert images. Delete the ``\vspace'' line,
% uncomment the preceding line ``\centerline...'' and replace ``imageX.ps''
% with a suitable PostScript file name.
% -------------------------------------------------------------------------


\bibliographystyle{IEEEbib}
\bibliography{strings,refs}

\end{document}